# Fabrication and Characterization of Flexible $In_2O_3$ Thin-Film Transistors on Mica Substrates

Hyeonjun Kong[1], Ping-Hsein Wu[2], Xian-Ye Jiang[2], Yasutaka Matsuo[3, 4],
Ying-Hao Chu[2,*], and Hiromichi Ohta[3, 4, *]

[1] *Graduate School of Information Science and Technology, Hokkaido University, N14W9, Kita, 060-0814 Sapporo, Japan*

[2] *Department of Materials Science and Engineering, National Tsing Hua University, Hsinchu 300044, Republic of China*

[3] *Research Institute for Electronic Science, Hokkaido University, N20W10, Kita, Sapporo 001-0020, Japan*

[4] *Institute for Frontier Education and Research on Semiconductors, Hokkaido University, N8W5, Sapporo 060-0808, Japan*

*Corresponding authors:
Ying-Hao Chu (yhchu@mx.nthu.edu.tw),
Hiromichi Ohta (hiromichi.ohta@es.hokudai.ac.jp)

**ABSTRACT**: Flexible oxide-semiconductor thin-film transistors (TFTs) with high field-effect mobility ($\mu_{FE}$ > 70 cm$^2$ V$^{-1}$ s$^{-1}$) are required for next-generation flexible displays. Polycrystalline $In_2O_3$-based TFTs are promising candidates; however, conventional flexible substrates such as plastics are not suitable because their fabrication typically requires relatively high temperatures (~350 °C). In this study, we used mica as a flexible substrate to fabricate polycrystalline $In_2O_3$-based TFTs, motivated by previous demonstrations of crystalline oxide growth on mica. The resulting TFTs exhibited a $\mu_{FE}$ of approximately 75 cm$^2$ V$^{-1}$ s$^{-1}$. The devices maintained gate-modulated operation after 10$^3$ flex-out bending cycles at a 5 mm radius, whereas flex-in cycling caused early electrical failure, likely due to $In_2O_3$ film degradation. The TFTs also operated normally after exposure to 85 °C and 40% relative humidity for one day. These results demonstrate that mica substrates enable high-mobility, flex-out-compatible $In_2O_3$-based TFTs.

## INTRODUCTION

Oxide semiconductor thin-film transistors (TFTs) are widely used as switching and driving devices in flat-panel displays.[1, 2] Their optical transparency, low off-state current, and large area uniformity make them attractive for flexible electronics.[1, 3, 4] Although flexible amorphous $InGaZnO_4$ (a-IGZO) TFTs have been demonstrated and their field-effect mobility ($\mu_{FE}$) is approximately 10 cm$^2$ V$^{-1}$ s$^{-1}$,[3, 5] higher pixel densities and refresh rates require sufficient drive current within a shorter pixel-charging time.[6] Thus, the development of high-mobility oxide channels has attracted significant attention.[7, 8] Among various oxide semiconductor TFTs, polycrystalline $In_2O_3$ TFTs are a promising candidate because of their high $\mu_{FE}$ (>70 cm$^2$ V$^{-1}$ s$^{-1}$).[9-12] For example, our previous study using an $In(OH)_3$ ceramic target yields a $\mu_{FE}$ of approximately 90 cm$^2$ V$^{-1}$ s$^{-1}$ on rigid substrates.[13]

For flexible a-IGZO TFTs, organic substrates such as polyethylene terephthalate (PET) have been used. Similarly, researchers have fabricated flexible $In_2O_3$ TFTs on polymer substrates using several fabrication methods.[14-18] The solution-processed flexible $In_2O_3$ TFTs using polymer substrates summarized in **Table 1** reported low $\mu_{FE}$ (0.24 to 14.5 cm$^2$ V$^{-1}$ s$^{-1}$) because the channel consisted of tiny grains; because polymer substrates are thermally unstable, the TFTs could not be annealed at high temperatures. Thus, flexible inorganic substrates with sufficient thermal stability[19, 20] are more suitable for realizing high-performance flexible $In_2O_3$ TFTs.

In this study, we chose mica as the substrate to realize flexible $In_2O_3$ TFTs. Mica combines flexibility with thermal stability required for high-temperature annealing. Its layered crystal structure allows cleavage into thin sheets with an atomically smooth surface.[21, 22]

Previous studies on crystalline oxide growth and flexible oxide devices on mica show that it can accommodate elevated-temperature processing.[22-24] In particular, flexible InGaZnO TFTs have been fabricated on mica, demonstrating its applicability as a substrate for oxide TFTs.[24] Therefore, mica is an ideal substrate for fabricating high-mobility polycrystalline $In_2O_3$ channels using a thermal annealing process.

Here, we fabricate $In_2O_3$ TFTs on flexible mica substrates and evaluate their electrical transfer performance, cyclic bending response, and operation after environmental exposure. The TFTs exhibit a $\mu_{FE}$ of approximately 75 $cm^2\ V^{-1}\ s^{-1}$, comparable to that of the rigid ITO-glass reference, and retain gate-modulated operation after $10^3$ outward bending cycles at a radius of 5 mm. We also evaluated their performance following an environmental exposure test at 85 °C and 40% relative humidity for one day.

## EXPERIMENTAL SECTION

**Fabrication of $In_2O_3$ TFTs**: Mica sheets (10 × 10 $mm^2$, SHINKOSHA CO., LTD.) were used. A 50-nm-thick Al-doped ZnO (AZO) buffer layer was deposited on mica, followed by a 200-nm-thick ITO layer. Both layers were deposited by RF sputtering at room temperature. The RF power was 200 W, and the Ar and $O_2$ flow rates were 30 and 1 sccm, respectively. The working pressure was 0.4 Pa. After deposition, the ITO gate electrode was annealed at 500 °C for 10 h in vacuum (6.7 Pa) to reduce the resistivity. A 100-nm-thick amorphous $Al_2O_3$ gate insulator was deposited by atomic layer deposition (ALD) at 300 °C. Next, a 5-nm-thick $In_2O_3$ channel layer was deposited at room temperature by pulsed laser deposition (PLD) through a stencil mask. An $In(OH)_3$ ceramic target was used. The oxygen partial pressure was kept at 2.5 Pa, and a KrF excimer laser ($\lambda$ = 248 nm) with a fluence of approximately 1.3 J $cm^{-2}$ was used. The multilayer film was annealed in air at 350 °C to crystallize the $In_2O_3$ channel. Finally, 100-nm-thick ITO source/drain electrodes were deposited by PLD through a stencil mask. The TFT channel width ($W$) and length ($L$) were 400 and 100 µm, respectively.

**Transistor characteristics measurements**: Transfer characteristics of the TFTs were measured using a semiconductor device analyzer (B1500A, Keysight) in ambient air in the dark. The drain voltage $V_d$ was 0.1 V, and the gate voltage $V_g$ was swept from −10 to +20 V in forward and reverse directions in steps of 0.1 V. Drain current ($I_d$) and gate leakage current ($I_g$) were recorded simultaneously. The linear-regime field-effect mobility was extracted from the transconductance using $\mu_{FE} = (L/W \cdot C_i \cdot V_d)(\partial I_d/\partial V_g)$, where $C_i$ is the gate-insulator capacitance per unit area and $\partial I_d/\partial V_g$ was obtained from the approximately linear portion of the $I_d$−$V_g$ curve.

**Bending tests**: The mechanical reliability of the $In_2O_3$ TFT on mica was evaluated by cyclic bending at a bending radius of 5 mm in flex-out and flex-in configurations, which induce tensile and compressive strain, respectively. Transfer characteristics were measured before bending and after 1, 10, $10^2$, and $10^3$ bending cycles.

**Environmental tests**: Environmental exposure was performed in an oven (DOV-450A, AS-ONE) set to 85 °C, with a container of water placed inside to provide moisture. Relative humidity was measured separately using an SHT-30 humidity sensor; the measured value was 40%. Transfer curves were measured before exposure and after exposure times of 4 h, 16 h, and 24 h. For side sealing, silicone liquid gasket (ThreeBond 1215) was applied thinly along all four edges of the mica samples and cured at room temperature for 24 h before environmental exposure.

## RESULTS AND DISCUSSION

Bottom-gate, top-contact $In_2O_3$ TFTs were fabricated on mica. After deposition, the $In_2O_3$ channel was thermally annealed at 350 °C to promote crystallization.[11, 13] **Figure 1a** shows the schematic device structure consisting of mica, an ITO bottom-gate electrode, a 100 nm-thick amorphous $Al_2O_3$ gate insulator, a 5 nm-thick $In_2O_3$ channel, and 100 nm-thick ITO source/drain electrodes. **Figure 1b** shows a device in a bent state without visible macroscopic fracture. The initial transfer characteristics show n-channel switching with low off-state current and small hysteresis (**Fig. 1c**). The TFTs on mica exhibit a $\mu_{FE}$ of approximately 75 cm$^2$ V$^{-1}$ s$^{-1}$, compared with approximately 86 cm$^2$ V$^{-1}$ s$^{-1}$ for the ITO-glass reference. For the TFTs on mica, the threshold voltage ($V_{th}$), subthreshold swing (SS), and on/off current ratio were −5.48 V, 0.19 V/decade, and 5.1 $\times 10^7$, respectively. The corresponding values for the TFTs on ITO glass were −5.3 V, 0.12 V/decade, and $3.1 \times 10^7$, respectively. **Figure 1d** shows the output characteristics of the $In_2O_3$ TFTs on mica and ITO glass, measured at gate voltages from 0 to 20 V in steps of 4 V. Both devices exhibit an approximately linear increase in $I_d$ at low $V_d$. At $V_g$ = 0 and 4 V, the curves tend to flatten as $V_d$ increases, whereas no clear saturation is observed at $V_g$ = 12 to 20 V within the measured $V_d$ range of 0 to 5 V. At $V_g$ = 20 V and $V_d$ = 5 V, the mica TFT exhibits an $I_d$ of 1.52 mA, compared with 1.67 mA for the ITO-glass reference, indicating comparable current-driving capability under these bias conditions.

We then performed bending tests as shown in **Fig. 2a** and measured the transfer characteristics after $10^0$, $10^1$, $10^2$, and $10^3$ bending cycles. The mechanical response to cyclic bending strongly depends on bending direction. At a bending radius $R$ of 5 mm, the strain at the mica surface was estimated to be approximately 0.2% using $\varepsilon = t/2R$, with a mica thickness t of approximately 20 μm.

In the flex-out configuration, the TFT retained gate-modulated operation after $10^3$ cycles, with closely overlapping transfer curves and only small changes in on-state drain current (**Fig. S1a**). We observed no catastrophic gate leakage under these conditions. Retained gate modulation therefore indicates continued switching operation, rather than unchanged electrical characteristics. On the other hand, the flex-in response was markedly different (**Fig. S1b**). After $10^0$ and $10^1$ cycles, the transfer curves shifted toward more negative gate voltages, although the TFT remained operational. After $10^2$ cycles, the gate current increased during the voltage sweep and eventually reached the compliance limit, indicating electrical breakdown. This behavior contrasts sharply with the low-leakage flex-out response, demonstrating that electrical failure under bending depends strongly on the bending direction.

**Figure 2b** summarizes how bending direction affects the transistor characteristics. The TFT retained gate-modulated operation, with only small changes in the threshold voltage, normalized on-current, and gate leakage current after $10^3$ repeated flex-out cycles. In contrast, after $10^2$ flex-in bending cycles, the threshold voltage shifted significantly. Flex-out and flex-in bending apply tensile and compressive deformation, respectively, to the device layers. Brittle films can respond differently to these deformation modes,[25-28] and direction-dependent degradation has been reported for flexible oxide TFTs.[29]

To clarify the origin of degradation during the flex-in bending test, we fabricated 30-nm-thick $In_2O_3$ model films on mica and performed a bending test similar to that for the TFTs. **Figure 3a** shows the change in electrical resistivity with the number of bending cycles. The resistivity increased significantly after flex-in bending, while the increase was milder after flex-out bending. We observed the surface morphology of the $In_2O_3$ film after the bending tests (**Fig. 3b**). After the flex-out bending test, we observed a rather smooth, dense surface composed of ~1 μm grains. In contrast, the flex-in bending test revealed a porous structure with ~1 μm pit-like surface features. These observations suggest a possible contribution from damage to $In_2O_3$ or its interfaces.

We then examined the environmental stability of the TFTs. We first exposed the TFTs to 85 °C and 40% relative humidity (**Fig. 4a**).[30] One day later, the TFT could not show transistor characteristics because intercalation of water molecules into the mica layer occurred. To prevent water-molecule intercalation,[31-34] we sealed the edge of the mica substrate using silicone liquid gasket (ThreeBond 1215) and performed the test in the same way (**Fig. 4b**). The initial, 4 h, 16 h, and 24 h drain-current curves remain close, although the gate current varies after exposure. These results show that sealing the mica edges enables the tested TFTs to operate under these environmental conditions.

The comparable $\mu_{FE}$ values (approximately 75 and 86 $cm^2\ V^{-1}\ s^{-1}$ for the mica and ITO-glass TFTs, respectively), $V_{th}$, and on/off current ratios demonstrate that the process incorporating crystallization at 350 °C can be applied to flexible mica substrates while retaining high mobility. However, the larger SS of the mica TFTs (0.19 vs 0.12 V/decade) indicates that comparable mobility does not necessarily imply equivalent subthreshold characteristics. The origin of this difference remains unclear. The failure behavior differed between flex-out and flex-in bending, even at the same bending radius and estimated surface strain magnitude. This suggests that bending radius or strain magnitude alone is insufficient to describe mechanical reliability; the responses of the constituent layers and interfaces to tensile and compressive deformation must also be considered. Buckling or interfacial delamination may contribute to damage under compression,[25, 35] although neither mechanism was directly confirmed in this study.

The more pronounced resistivity increase and pit-like surface features in the 30-nm-thick $In_2O_3$ model films after flex-in bending are consistent with a possible contribution from $In_2O_3$-related damage to the direction-dependent degradation. However, the model films differ from the 5-nm-thick TFT channels in both thickness and layer structure. Furthermore, the resistivity increase alone does not directly explain the negative $V_{\mathrm{th}}$ shift or the large gate leakage current observed in the TFTs. The increased gate leakage points to the formation of an abnormal leakage path between the gate and the source/drain terminals, but its location and origin remain unidentified. Thus, the model-film results do not establish the failure mechanism in the complete TFT structure.

Under the same exposure conditions, the unsealed mica TFT lost gate-modulated operation (**Fig. 4a**), whereas the edge-sealed mica TFT remained operational (**Fig. 4b**). This contrast suggests that factors associated with the mica substrate or the edges of the mica-based layer stack contribute to the degradation. The effectiveness of edge sealing is consistent with moisture ingress through the edges.[30, 36] These results suggest that the thermal stability required for fabrication does not, by itself, ensure environmental durability after fabrication. Edge sealing effectively maintained gate-modulated operation under the conditions examined here. However, gate current changes were observed even in the sealed TFTs, and testing was limited to a single temperature–humidity condition for up to 24 h.

Future work should investigate structures that reduce compressive strain in the active layer, for example by adjusting the neutral-plane position or reducing the mica thickness,[37, 38] and examine failed regions to identify the leakage path. The material and geometry of the edge sealing should also be optimized, and its protective effect should be evaluated under longer environmental exposure and after repeated bending.

## CONCLUSION

We fabricated flexible $In_2O_3$ TFTs on a mica substrate and evaluated their electrical transfer performance, cyclic bending response, and operation after environmental exposure. The TFTs exhibit a $\mu_{FE}$ of approximately 75 $cm^2 V^{-1} s^{-1}$, comparable to the rigid ITO-glass reference, and retain gate-modulated operation after $10^3$ outward bending cycles at a 5 mm radius. We also evaluated their performance following an environmental exposure test at 85 °C and 40% relative humidity for one day. These results show that mica enables high-mobility flexible $In_2O_3$ TFTs that tolerate repeated flex-out bending and remain environmentally stable with side-sealing.

## AUTHOR INFORMATION

### Corresponding Authors

Ying-Hao Chu − Department of Materials Science and Engineering, National Tsing Hua University, Hsinchu 300044, Republic of China

ORCiD: orcid.org/ 0000-0002-3435-9084

Email: yhchu@mx.nthu.edu.tw

Hiromichi Ohta − Research Institute for Electronic Science, Hokkaido University, N20W10, Kita, Sapporo 001-0020, Japan

ORCiD: orcid.org/0000-0001-7013-0343

Email: hiromichi.ohta@es.hokudai.ac.jp

### Author

Hyeonjun Kong – Graduate School of Information Science and Technology, Hokkaido University, N14W9, Kita, Sapporo 060-0814, Japan

ORCiD: orcid.org/0009-0004-0358-2162

Ping-Hsein Wu − Department of Materials Science and Engineering, National Tsing

Hua University, Hsinchu 300044, Republic of China
ORCiD: orcid.org/0009-0004-6578-6785

Xian-Ye Jiang − Department of Materials Science and Engineering, National Tsing Hua University, Hsinchu 300044, Republic of China
ORCiD: orcid.org/0009-0005-9988-1833

Yasutaka Matsuo − Research Institute for Electronic Science, Hokkaido University, N20W10, Kita, Sapporo 001-0020, Japan
ORCiD: orcid.org/0000-0002-5071-0284

**Author Contributions**

H.K. fabricated the samples and performed the measurements. Y.H.C and X.Y.J. provided the use of mica substrates. H.O. supervised the project. All authors discussed the results and commented on the manuscript.

**Funding Sources**

This work was supported by the Japan Society for the Promotion of Science (JSPS) KAKENHI Grant Number 26H02188 (HO). H.K. was supported by JST SPRING (JPMJSP2119). Part of this work was supported by the Crossover Alliance to Create the Future with People, Intelligence and Materials, and by the Network Joint Research Center for Materials and Devices.

**Notes**

The authors declare no competing interests.

**ACKNOWLEDGEMENTS**

This work was supported by the Japan Society for the Promotion of Science (JSPS) KAKENHI Grant Number 26H02188 (HO). H.K. was supported by JST SPRING (JPMJSP2119). This work was partly supported by the Crossover Alliance to Create the Future with People, Intelligence and Materials, and the Network Joint Research Center for Materials and Devices.

**REFERENCES**

(1) Kamiya, T.; Nomura, K.; Hosono, H. Present status of amorphous In-Ga-Zn-O thin-film transistors. *Sci. Technol. Adv. Mater.* **2010**, *11* (4), 044305. DOI: 10.1088/1468-6996/11/4/044305.

(2) Wu, Z.; Yuan, G.; Xie, H.; Cao, K.; Li, Y.; Zeng, S.; Zhou, Y.; Yang, F.; Cheng, J.; Shi, H.; et al. 64.3: development of 55-inch UHD AMOLED TV. *SID Symp. Dig. Tech. Pap.* **2015**, *46* (1), 954–957. DOI: 10.1002/sdtp.10427.

(3) Nomura, K.; Ohta, H.; Takagi, A.; Kamiya, T.; Hirano, M.; Hosono, H. Room-temperature fabrication of transparent flexible thin-film transistors using amorphous oxide semiconductors. *Nature* **2004**, *432* (7016), 488–492. DOI: 10.1038/nature03090.

(4) Sheng, J.; Jeong, H.-J.; Han, K.-L.; Hong, T.; Park, J.-S. Review of recent advances in flexible oxide semiconductor thin-film transistors. *J. Inf. Disp.* **2017**, *18* (4), 159–172. DOI: 10.1080/15980316.2017.1385544.

(5) Sato, A.; Shimada, M.; Abe, K.; Hayashi, R.; Kumomi, H.; Nomura, K.; Kamiya, T.; Hirano, M.; Hosono, H. Amorphous In-Ga-Zn-O thin-film transistor with coplanar homojunction structure. *Thin Solid Films* **2009**, *518* (4), 1309–1313. DOI: 10.1016/j.tsf.2009.01.165.

(6) Liao, Y. P.; Shao, X. B.; Du, Y. F.; Song, Y. Z.; Hu, W. H.; Zhang, Z. Y.; Chen, Y. Q.; Wang, Y.; Ma, Q.; Yoon, D.; et al. Development of a 120 Hz 110" ultra-high-definition a-Si liquid crystal display panel. *J. Inf. Disp.* **2014**, *15* (2), 77–80. DOI: 10.1080/15980316.2014.907214.

(7) Yang, H.; Zhang, Y.; Matsuo, Y.; Magari, Y.; Ohta, H. Thermopower Modulation Analyses of High-Mobility Transparent Amorphous Oxide Semiconductor Thin-Film Transistors. *ACS Appl. Electron. Mater.* **2022**, *4* (10), 5081–5086. DOI: 10.1021/acsaelm.2c01210.

(8) Park, J.-S. Oxide TFTs for AMOLED TVs. *Inf. Disp.* **2013**, *29* (2), 16–19. DOI: 10.1002/j.2637-496X.2013.tb00593.x.

(9) Körber, C.; Krishnakumar, V.; Klein, A.; Panaccione, G.; Torelli, P.; Walsh, A.; Da Silva, J. L. F.; Wei, S.-H.; Egdell, R. G.; Payne, D. J. Electronic structure of $In_2O_3$ and Sn-doped $In_2O_3$ by hard x-ray photoemission spectroscopy. *Phys. Rev. B* **2010**, *81* (16), 165207. DOI: 10.1103/PhysRevB.81.165207.

(10) Koida, T.; Fujiwara, H.; Kondo, M. Hydrogen-doped $In_2O_3$ as high-mobility transparent conductive oxide. *Jpn. J. Appl. Phys.* **2007**, *46* (25–28), L685–L687. DOI:

10.1143/JJAP.46.L685.
(11) Magari, Y.; Kataoka, T.; Yeh, W.; Furuta, M. High-mobility hydrogenated polycrystalline $In_2O_3$ ($In_2O_3$:H) thin-film transistors. *Nat. Commun.* **2022**, *13* (1), 1078. DOI: 10.1038/s41467-022-28480-9.
(12) Ghediya, P. R.; Magari, Y.; Sadahira, H.; Endo, T.; Furuta, M.; Zhang, Y.; Matsuo, Y.; Ohta, H. Reliable Operation in High-Mobility Indium Oxide Thin Film Transistors. *Small Methods* **2025**, *9* (1), e2400578. DOI: 10.1002/smtd.202400578.
(13) Sadahira, H.; Ghediya, P. R.; Kong, H.; Miura, A.; Matsuo, Y.; Ohta, H.; Magari, Y. Indium Hydroxide Ceramic Targets: A Breakthrough in High-Mobility Thin-Film Transistor Technology. *ACS Appl. Electron. Mater.* **2025**, *7* (15), 6952–6959. DOI: 10.1021/acsaelm.5c00829.
(14) Bhalerao, S. R.; Lupo, D.; Berger, P. R. Flexible, solution-processed, indium oxide ($In_2O_3$) thin film transistors (TFT) and circuits for internet-of-things (IoT). *Mater. Sci. Semicond. Process.* **2022**, *139*, 106354. DOI: 10.1016/j.mssp.2021.106354.
(15) Leppäniemi, J.; Huttunen, O. H.; Majumdar, H.; Alastalo, A. Flexography-printed $In_2O_3$ semiconductor layers for high-mobility thin-film transistors on flexible plastic substrate. *Adv. Mater.* **2015**, *27* (44), 7168–7175. DOI: 10.1002/adma.201502569.
(16) Hwang, Y. H.; Seo, J.-S.; Yun, J. M.; Park, H. J.; Yang, S.; Park, S.-H. K.; Bae, B.-S. An 'aqueous route' for the fabrication of low-temperature-processable oxide flexible transparent thin-film transistors on plastic substrates. *NPG Asia Mater.* **2013**, *5* (4), e45. DOI: 10.1038/am.2013.11.
(17) Leppäniemi, J.; Eiroma, K.; Majumdar, H.; Alastalo, A. Far-UV Annealed Inkjet-Printed $In_2O_3$ Semiconductor Layers for Thin-Film Transistors on a Flexible Polyethylene Naphthalate Substrate. *ACS Appl. Mater. Interfaces* **2017**, *9* (10), 8774–8782. DOI: 10.1021/acsami.6b14654.
(18) Tewari, A.; Tukiainen, A.; Mäntysalo, M.; Berger, P. R. Direct PEALD Deposition of a $HfO_2$ Gate Dielectric without the Passivation for TFTs on Rigid and Flexible Substrates. *ACS Appl. Electron. Mater.* **2025**, *7* (15), 7120–7130. DOI: 10.1021/acsaelm.5c00953.
(19) MacDonald, W. A. Engineered films for display technologies. *J. Mater. Chem.* **2004**, *14* (1), 4–10. DOI: 10.1039/b310846p.
(20) Han, K.-L.; Jeong, H.-J.; Kim, B.-S.; Lee, W.-B.; Park, J.-S. Recent review on improving mechanical durability for flexible oxide thin film transistors. *J. Phys. D: Appl. Phys.* **2019**, *52* (48), 483002. DOI: 10.1088/1361-6463/ab3b6b.
(21) Bitla, Y.; Chu, Y.-H. MICAtronics: A new platform for flexible X-tronics.

*FlatChem* **2017**, *3*, 26–42. DOI: 10.1016/j.flatc.2017.06.003.
(22) Bitla, Y.; Chen, C.; Lee, H.-C.; Do, T. H.; Ma, C.-H.; Qui, L. V.; Huang, C.-W.; Wu, W.-W.; Chang, L.; Chiu, P.-W.; et al. Oxide Heteroepitaxy for Flexible Optoelectronics. *ACS Appl. Mater. Interfaces* **2016**, *8* (47), 32401–32407. DOI: 10.1021/acsami.6b10631.
(23) Jiang, J.; Bitla, Y.; Huang, C.-W.; Do, T. H.; Liu, H.-J.; Hsieh, Y.-H.; Ma, C.-H.; Jang, C.-Y.; Lai, Y.-H.; Chiu, P.-W.; et al. Flexible ferroelectric element based on van der Waals heteroepitaxy. *Sci. Adv.* **2017**, *3* (6), e1700121. DOI: 10.1126/sciadv.1700121.
(24) Huo, W. X.; Mei, Z. X.; Sui, Y. X.; Han, Z. Y.; Wang, T.; Liang, H. L.; Du, X. L. Flexible Transparent InGaZnO Thin-Film Transistors on Muscovite Mica. *IEEE Trans. Electron Devices* **2019**, *66* (5), 2198–2201. DOI: 10.1109/TED.2019.2902346.
(25) Chen, Z.; Cotterell, B.; Wang, W. The fracture of brittle thin films on compliant substrates in flexible displays. *Eng. Fract. Mech.* **2002**, *69* (5), 597–603. DOI: 10.1016/S0013-7944(01)00104-7.
(26) Méndez-Hernández, J.; Hernández-Pérez, A.; Oviedo-Mendoza, M.; Hernández-Rodríguez, E. Effects of mechanical deformations on P3HT:PCBM layers for flexible solar cells. *Mech. Mater.* **2021**, *154*, 103708. DOI: 10.1016/j.mechmat.2020.103708.
(27) Wang, Y.; Li, M.; Fan, B.; Wong, Y. S.; Lo, C. Y.; Kwok, C. K. G.; Shil, S. K.; Yip, H.-L.; Jen, A. K.-Y.; Tsang, S.-W.; et al. Flexibility of Room-Temperature-Synthesized Amorphous CdO-$In_2O_3$ Alloy Films and Their Application as Transparent Conductors in Solar Cells. *ACS Appl. Mater. Interfaces* **2021**, *13* (36), 43795–43805. DOI: 10.1021/acsami.1c14722.
(28) Mohammed, D. W. Mechanical flexibility and electrical reliability of ZnO-Al thin films on polymer substrates under different external deformation. *Adv. Mater. Sci.* **2024**, *24* (1), 32–45. DOI: 10.2478/adms-2024-0003.
(29) Billah, M. M.; Hasan, M. M.; Jang, J. Effect of Tensile and Compressive Bending Stress on Electrical Performance of Flexible a-IGZO TFTs. *IEEE Electron Device Lett.* **2017**, *38* (7), 890–893. DOI: 10.1109/LED.2017.2707279.
(30) Wu, K.; Mariello, M.; Leterrier, Y.; Lacour, S. P. Optical Monitoring of Water Side Permeation in Thin Film Encapsulation. *Adv. Mater.* **2024**, *36* (24), e2310201. DOI: 10.1002/adma.202310201.
(31) Dollekamp, E.; Bampoulis, P.; Faasen, D. P.; Zandvliet, H. J. W.; Kooij, E. S. Charge Induced Dynamics of Water in a Graphene-Mica Slit Pore. *Langmuir* **2017**, *33* (43), 11977–11985. DOI: 10.1021/acs.langmuir.7b02759.
(32) Rauf, A.; Cojal González, J. D.; Balkan, A.; Severin, N.; Sokolov, I. M.; Rabe, J. P.

Shaping surfaces and interfaces of 2D materials on mica with intercalating water and ethanol. *Mol. Phys.* **2021**, *119* (15−16), e1947534. DOI: 10.1080/00268976.2021.1947534.
(33) Ochedowski, O.; Bussmann, B. K.; Schleberger, M. Graphene on Mica - Intercalated Water Trapped for Life. *Sci. Rep.* **2014**, *4* (1), 6003. DOI: 10.1038/srep06003.
(34) Severin, N.; Lange, P.; Sokolov, I. M.; Rabe, J. P. Reversible dewetting of a molecularly thin fluid water film in a soft graphene-mica slit pore. *Nano Lett.* **2012**, *12* (2), 774–779. DOI: 10.1021/nl2037358.
(35) Yu, H.-H.; Hutchinson, J. W. Influence of substrate compliance on buckling delamination of thin films. *Int. J. Fract.* **2002**, *113* (1), 39–55. DOI: 10.1023/A:1013790232359.
(36) Kempe, M. D.; Dameron, A. A.; Reese, M. O. Evaluation of moisture ingress from the perimeter of photovoltaic modules. *Prog. Photovolt.: Res. Appl.* **2014**, *22* (11), 1159–1171. DOI: 10.1002/pip.2374.
(37) Sekitani, T.; Zschieschang, U.; Klauk, H.; Someya, T. Flexible organic transistors and circuits with extreme bending stability. *Nat. Mater.* **2010**, *9* (12), 1015–1022. DOI: 10.1038/nmat2896.
(38) Salvatore, G. A.; Münzenrieder, N.; Kinkeldei, T.; Petti, L.; Zysset, C.; Strebel, I.; Büthe, L.; Tröster, G. Wafer-scale design of lightweight and transparent electronics that wraps around hairs. *Nat. Commun.* **2014**, *5* (1), 2982. DOI: 10.1038/ncomms3982.

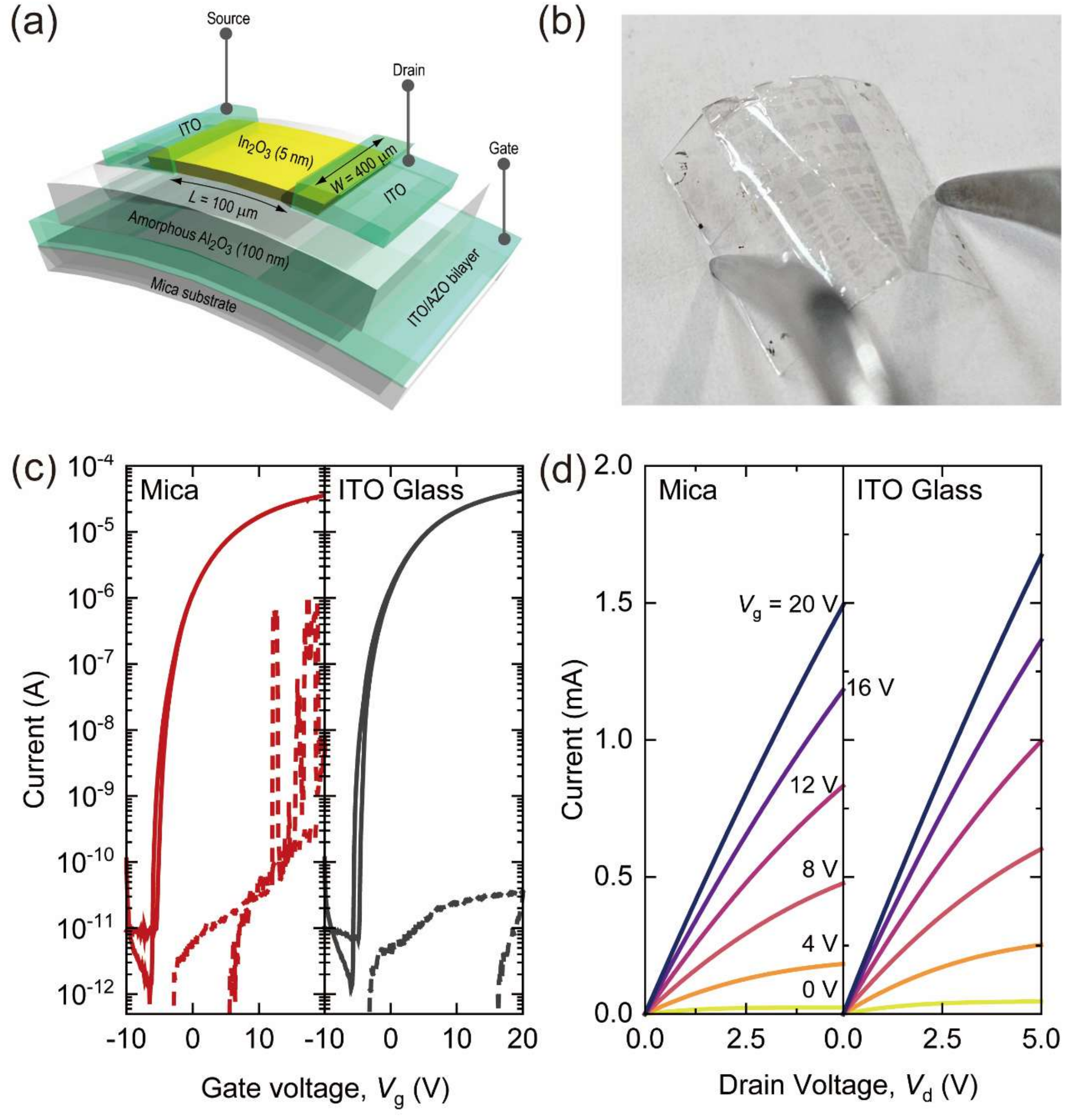

**Figure 1. Structure and electrical characteristics of $In_2O_3$ thin-film transistors (TFTs).** (a) Schematic of the bottom-gate, top-contact TFT on mica comprising an AZO buffer layer (50 nm), an ITO gate electrode (200 nm), an amorphous $Al_2O_3$ gate insulator (100 nm), an $In_2O_3$ channel (5 nm), and ITO source and drain electrodes (100 nm). The channel width and length are 400 and 100 μm, respectively. (b) Photograph of the device in a bent configuration. (c) Transfer characteristics measured during forward and reverse gate-voltage sweeps at $V_d = 0.1$ V. Solid and dashed curves denote drain and gate leakage currents, respectively. (d) Output characteristics measured over $V_d = 0$ to 5 V at $V_g = 0$ to 20 V.

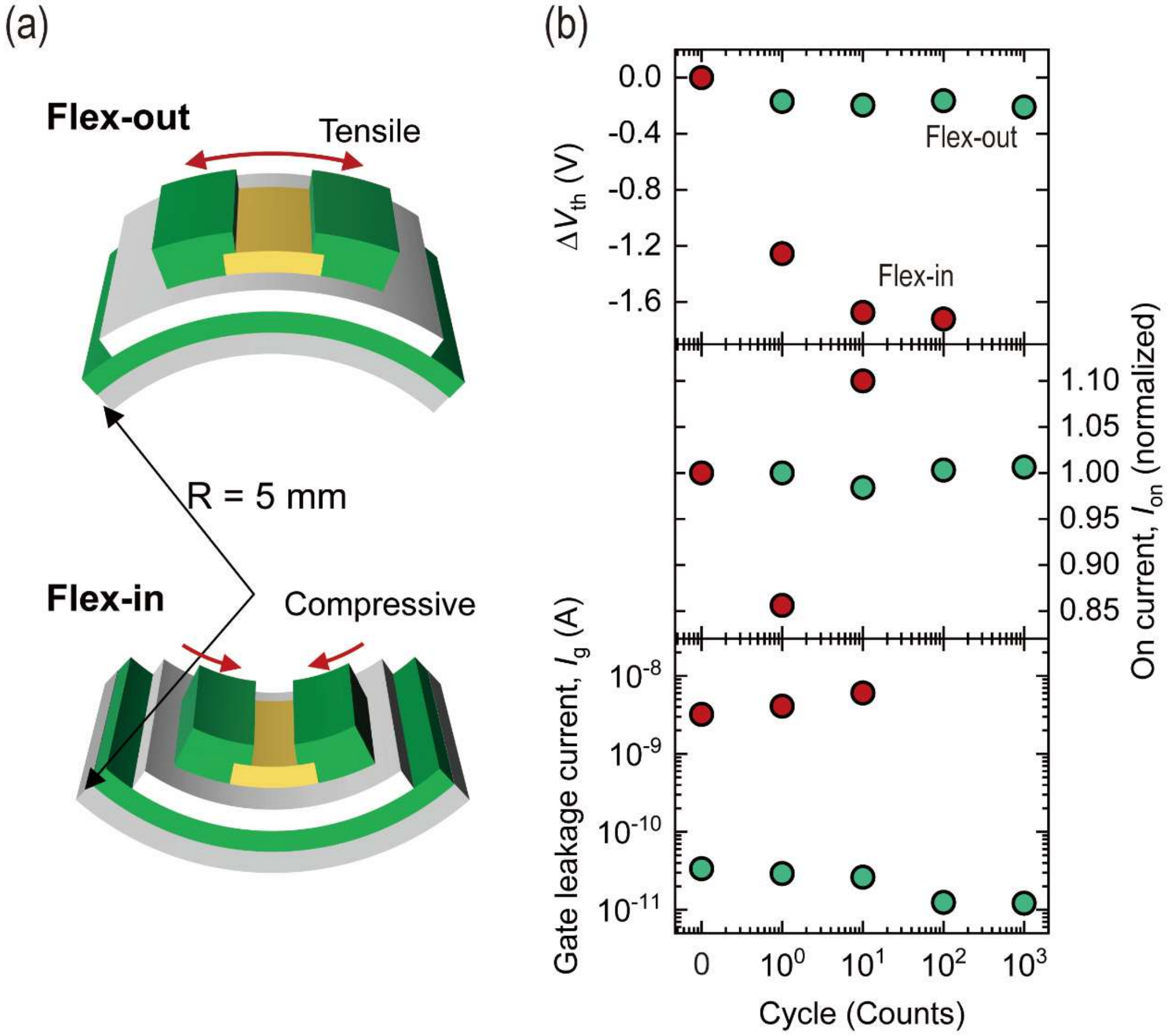

**Figure 2. Bending reliability of $In_2O_3$ TFTs on mica.** (a) Schematic of flex-out and flex-in bending at a radius of 5 mm, which apply tensile and compressive strain to the device layers, respectively. (b) Changes in threshold voltage, on-state drain current, and gate leakage current with bending cycle number. Gate-modulated operation is retained after $10^3$ flex-out cycles, whereas electrical breakdown occurs after $10^2$ flex-in cycles.

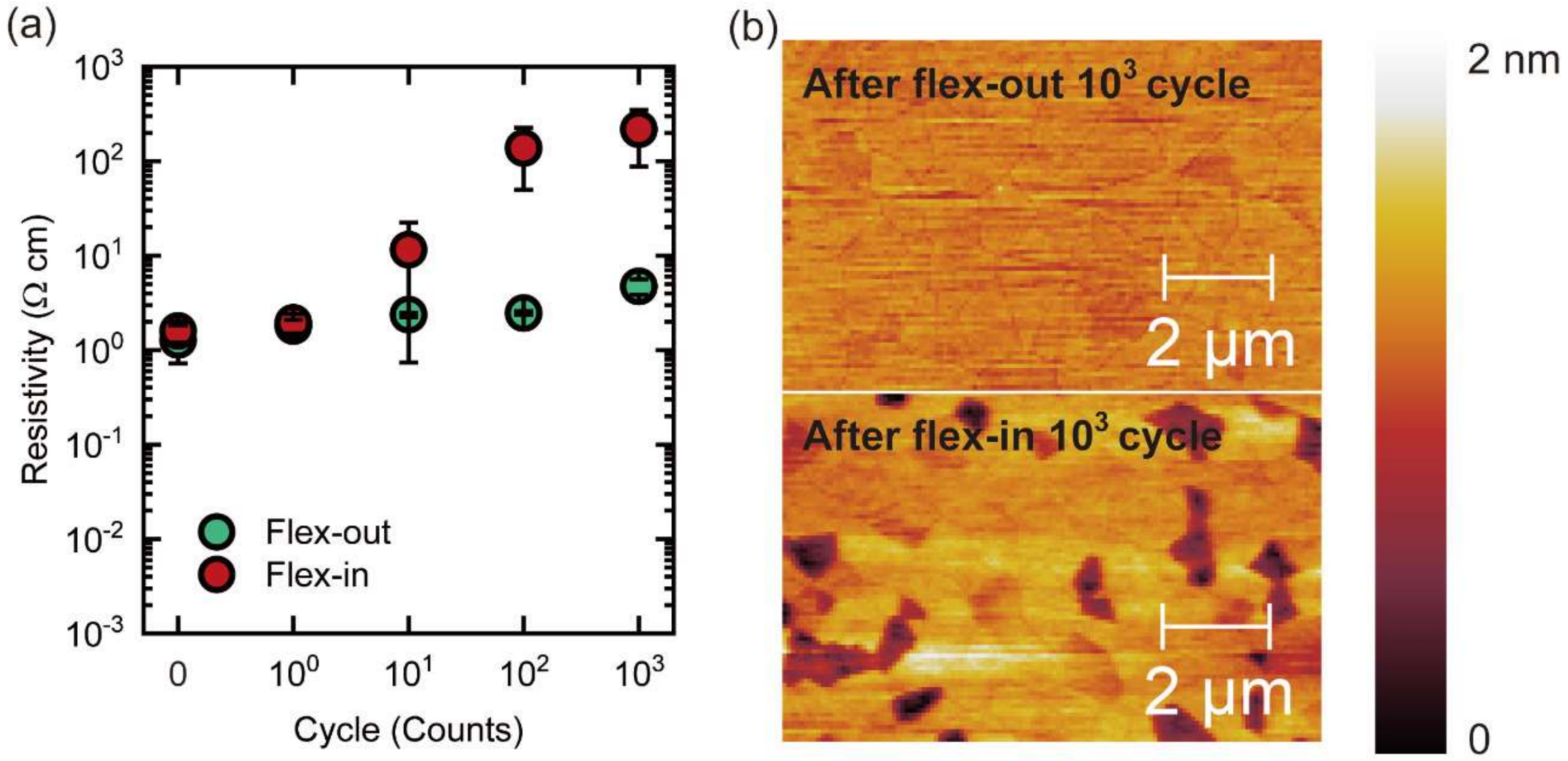

**Figure 3. Bending-induced changes in the electrical resistivity and surface morphology of 30-nm-thick $In_2O_3$ films on mica.** (a) Resistivity as a function of bending cycle number for flex-out and flex-in bending at a radius of 5 mm. Error bars represent the standard deviations of measurements taken at least five locations. (b) AFM images after $10^3$ cycles of flex-out (top) and flex-in (bottom) bending.

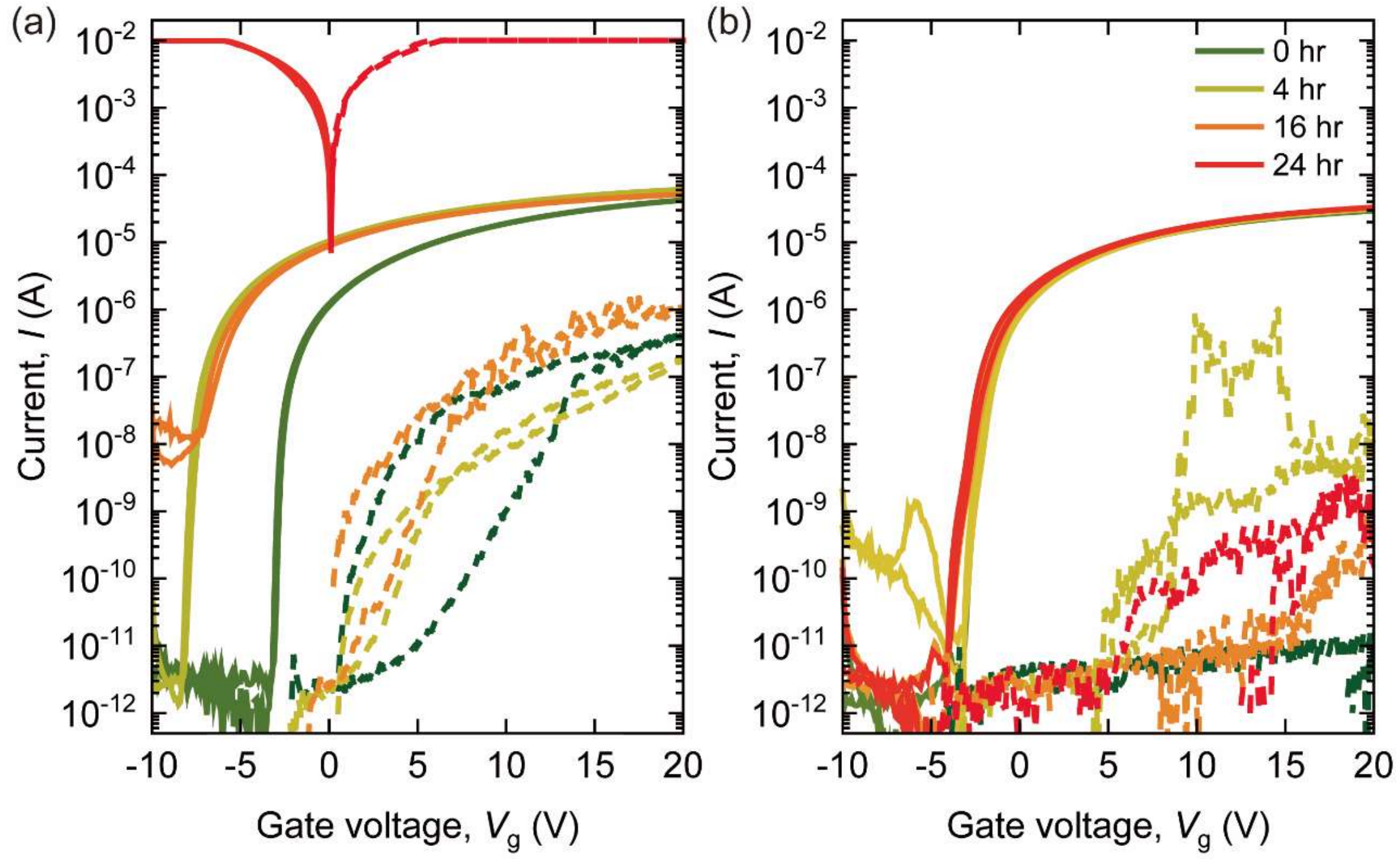

**Figure 4. Environmental stability of $In_2O_3$ TFTs on mica.** Transfer characteristics of $In_2O_3$ TFTs on (a) unsealed mica and (b) side-sealed mica before exposure and after 4, 16, and 24 h of exposure to an oven setting of 85 °C and a measured relative humidity of 40%. ThreeBond 1215 was applied along all four mica edges and cured at room temperature for 24 h before exposure. After 24 h, the unsealed TFT exhibits electrical breakdown, whereas the side-sealed TFT retains gate-modulated operation. Solid and dashed curves denote drain and gate leakage currents, respectively.

**Table 1. Comparison of fabrication conditions, and electrical characteristics of previously reported flexible $In_2O_3$ TFTs, InGaZnO TFTs on mica, and $In_2O_3$ TFTs on mica demonstrated in this work.** Mobilities are reported given in the original studies. A dash denotes a value not verified for the corresponding device. Processing temperature denotes the highest identified temperature during device fabrication. In this work, the AZO/ITO gate electrode was annealed at 500 °C, whereas the $In_2O_3$ channel was annealed at 350 °C

| **Ref.** | **Channel / Substrates** | **Processing temperature (°C)** | **$\mu_{FE}$ ($cm^2$ $V^{-1}$ $s^{-1}$)** | **SS (V dec$^{-1}$)** | **on/off ratio** |
|---|---|---|---|---|---|
| **14** | $In_2O_3$ / PI | 300 | 14.5 | 0.22 | $10^5$ |
| **15** | $In_2O_3$ / PI | 300 | 8 ± 4 | 0.4 ± 0.1 | > $10^6$ |
| **16** | $In_2O_3$ / PEN | 200 | 3.14 | 0.158 | > $10^9$ |
| **17** | $In_2O_3$ / PEN | 150 | 1 | − | − |
| **18** | $In_2O_3$ / PI | 250 | 0.24 ± 0.04 | 0.96 | $10^4$ |
| **24** | InGaZnO / Mica | 400 | 3.03 | 0.65 | $10^8$ |
| **This work** | $In_2O_3$ / Mica | 500 | ~ 75 | 0.19 | 5.1 × $10^7$ |

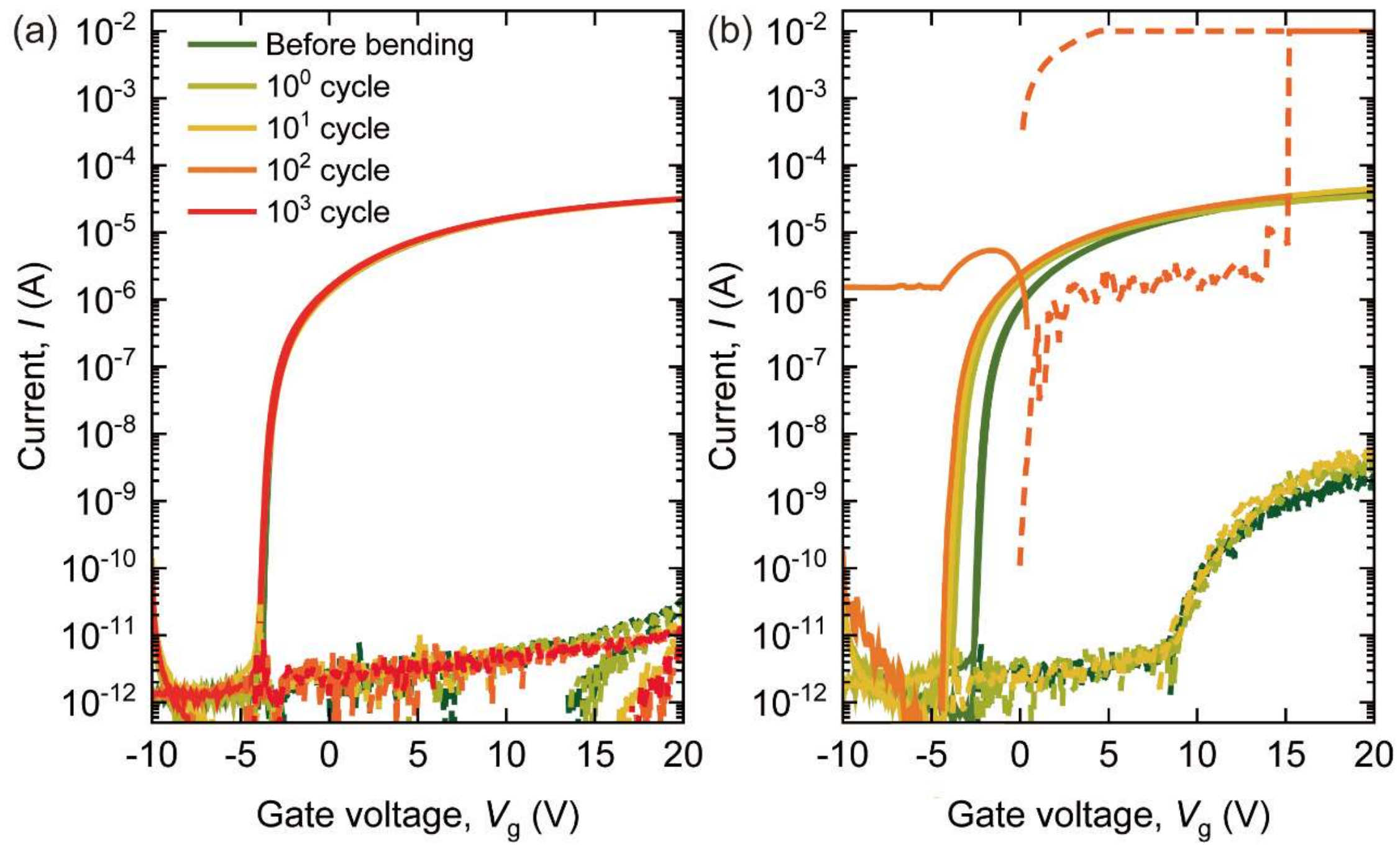

**Figure S1. Transfer characteristics of $In_2O_3$ TFTs on mica under cyclic bending.** Curves measured before and after repeated (a) flex-out and (b) flex-in bending at a radius of 5 mm. Measurements were taken in the initial state and after $10^0$, $10^1$, $10^2$, and $10^3$ cycles for flex-out bending, and after $10^0$, $10^1$, and $10^2$ cycles for flex-in bending. Gate-modulated operation is retained after $10^3$ flex-out cycles, whereas the gate current reaches the 10 mA compliance limit after $10^2$ flex-in cycles, indicating electrical breakdown. Solid and dashed curves denote drain and gate leakage currents, respectively.